\documentclass[aps,prc,superscriptaddress,showpacs,nofootinbib]{revtex4}
\usepackage{amsmath}

\newcommand{\mean}[1]{\left\langle #1 \right\rangle}
\newcommand{\smean}[1]{\langle #1 \rangle}

\newcommand{\Np}{N'}

\begin{document}

\preprint{Saclay-T04/056}

\title{Directed flow at RHIC from Lee--Yang zeroes}

\author{N. Borghini}
\email{borghini@spht.saclay.cea.fr}

\author{J.-Y. Ollitrault}
\email{Ollitrault@cea.fr}
\altaffiliation[also at ]{L.P.N.H.E., Universit{\'e} Pierre et Marie Curie,
4 place Jussieu, F-75252 Paris cedex 05, France}
\affiliation{Service de Physique Th{\'e}orique, CEA-Saclay,
F-91191 Gif-sur-Yvette cedex, France}

\date{\today}

\begin{abstract}
Directed flow in ultrarelativistic nucleus-nucleus collisions 
is analyzed using the event plane from elliptic flow, 
which reduces the bias from nonflow effects. 
We combine this method with the determination of elliptic 
flow from Lee--Yang zeroes. 
The resulting method is more consistent and somewhat 
easier to implement than the previously used method based 
on three-particle cumulants, and is also less biased by 
nonflow correlations. Error terms from residual nonflow 
correlations are carefully estimated, as well as statistical
errors. We discuss the application of the method at 
RHIC and LHC. 
\end{abstract}


\maketitle
\section{Introduction}

The standard method of analyzing directed flow ($v_1$) in
nucleus-nucleus collisions using an 
estimate of the reaction plane~\cite{Danielewicz:hn,Poskanzer:1998yz}
was shown to be inadequate at ultrarelativistic 
energies~\cite{Borghini:2000cm}, due to the smallness of 
directed flow and the relatively large magnitude of 
``nonflow'' effects~\cite{Poskanzer:1998yz,Ollitrault:dy}. 
Trivial nonflow effects such as momentum conservation can 
be taken into account in the analysis \cite{Borghini:2000cm}, 
and the standard analysis can be modified for this 
purpose~\cite{Borghini:2002mv}. 
A more systematic, model-independent way of eliminating 
nonflow effects was introduced~\cite{Borghini:2002vp},
based on the observation that the three-particle average 
$\smean{\cos(\phi_1+\phi_2-2\phi_3)}$ (where 
$\phi_1$, $\phi_2$ and $\phi_3$ denote the azimuthal 
angles of three particles emitted in a collision, 
and $\mean{\cdots}$ denotes an average over triplets of 
particles and events) is much less sensitive 
to nonflow effects than the two-particle correlation 
$\smean{\cos(\phi_1-\phi_2)}$ used in the standard analysis. 
Since $\smean{\cos(\phi_1+\phi_2-2\phi_3)}\propto v_1^2 v_2$, 
where the elliptic flow $v_2$ is large at ultrarelativistic 
energies, this provides an alternative way of analyzing 
directed flow. 
This method was first implemented at the CERN 
SPS by the NA49 Collaboration~\cite{Borghini:2002hm,Alt:2003ab}, 
and recently led to the discovery of directed flow at 
RHIC~\cite{Adams:2003zg}.
Directed flow at ultrarelativistic energies is interesting 
in itself~\cite{Snellings:1999bt}; in addition, it is the 
only way of measuring the sign of elliptic flow, and to check
experimentally that it is positive~\cite{Ollitrault:bk}.

The practical implementation of the three-particle method presented 
in Ref.~\cite{Borghini:2002vp} is rather cumbersome and requires
to estimate elliptic flow $v_2$ independently, using 
cumulants~\cite{Borghini:2000sa} to avoid nonflow effects.
In this paper, we suggest to analyze simultaneously elliptic 
and directed flows using the recently introduced method of 
Lee--Yang zeroes~\cite{Bhalerao:2003yq,Borghini:2004ke} to analyze 
elliptic flow. 
This minimizes the bias from nonflow effects for both $v_1$ and $v_2$. 
The practical recipe is presented in Sec.~\ref{s:implementation}. 
The theoretical background is briefly discussed in Sec.~\ref{s:theory}. 
In Sec.~\ref{s:errors},
we derive the general order of magnitude of the systematic 
error due to nonflow correlations, which was underestimated 
in Ref.~\cite{Borghini:2002vp}. We also give analytical expressions 
of statistical errors.

\section{Implementation}
\label{s:implementation}

Let us first define useful quantities and notations. 
For a given event, we define the following complex-valued function 
\begin{equation}
\label{defgenfunc}
g_\epsilon^{\theta_1,\theta_2}(z)=\prod_{j=1}^M\left[1
+z \epsilon w_{1}(j)\cos(\phi_j-\theta_1)
+z w_{2}(j)\cos(2(\phi_j-\theta_2))\right]
\end{equation}
where $\theta_1$ and $\theta_2$ are angles, $\epsilon$ is a 
real parameter, $z$ is a complex variable, $\phi_j$ are 
the azimuthal angles of the particles 
(measured using a fixed reference in the laboratory), 
and the product runs over all detected particles. 
As in other methods of analysis, $w_1$ and $w_2$ are weights appropriate 
to directed and elliptic flows, respectively, and can be any functions 
of particle type, transverse momentum $p_T$ and rapidity $y$. 
In Eq.~(\ref{defgenfunc}), $w_n(j)$ is a shorthand for $w_n({p_T}_j,y_j)$. 
The best weight is the flow itself~\cite{Danielewicz:1995,Borghini:2000sa}, 
$w_n(p_T,y) = v_n(p_T,y)$, where $v_n(p_t,y)$ denotes the value of the 
flow in a small $(p_T,y)$ bin. 
In practice, one can choose as a first guess the center-of-mass 
rapidity for directed flow, $w_1=y-y_{\rm CM}$, 
and the transverse momentum for elliptic flow $w_2=p_T$, 
in regions of phase space covered by the detector acceptance. 
The average of $g^{\theta_1,\theta_2}_\epsilon(z)$ 
over events in a centrality bin will be denoted 
by $\smean{g^{\theta_1,\theta_2}_\epsilon(z)}_{\rm evts}$. 
The same notation holds for any observable associated with 
the event.

The implementation of the method consists of three logical steps,
as in the 3-particle correlation method~\cite{Borghini:2002vp}. 
The first one is to analyze {\it integrated\/} elliptic flow
$V_2$. By ``integrated,'' we mean that it is 
summed over all detected particles with appropriate 
weights, and averaged over many events: 
\begin{equation}
\label{defV2}
V_2\equiv\mean{\sum_{j=1}^M w_2(j)\cos(2(\phi_j-\Phi_R))}_{\!\!\!\rm evts},
\end{equation}
where the sum runs over all particles detected in an event and
$\Phi_R$ is the azimuthal angle of the reaction plane 
of the event.\footnote{$\Phi_R$ is the {\it exact\/} reaction 
plane, not an estimated plane; $\Phi_R$ is unknown on an event-by-event
basis.}
The second step is the analysis of integrated directed flow 
$V_1$, which is defined similarly:
\begin{equation}
\label{defV1}
V_1\equiv\mean{\sum_{j=1}^M w_1(j)\cos(\phi_j-\Phi_R)}_{\!\!\!\rm evts}.
\end{equation}
Finally, in a third step, integrated values are used as a reference
to analyze directed flow differentially, as a function of 
transverse momentum and rapidity. 
In practice, the method requires only two passes through the data: 
the first pass corresponds to Sec.~\ref{s:IIA}, while 
the computations in Sec.~\ref{s:IIB} and Sec.~\ref{s:IIC} 
can be done in a single second pass. 

\subsection{Integrated elliptic flow}
\label{s:IIA}

The analysis of integrated elliptic flow is identical to that presented in
Refs.~\cite{Bhalerao:2003yq,Borghini:2004ke}, to which we refer the 
reader for further practical details. 
Note, however, that it uses the generating function introduced in 
Ref.~\cite{Borghini:2004ke}, not that of Ref.~\cite{Bhalerao:2003yq}. 
Let us just give a short reminder of the recipe. 

One must evaluate 
$g_\epsilon ^{\theta_1,\theta_2}(z)$ for $\epsilon=\theta_1=0$, 
for several (typically 4 or 5) equally spaced values of 
$\theta_2$ between $0$ and $\pi/2$ (i.e., $\theta_2=0,\pi/(2n),\ldots, 
(n-1)\pi/(2n)$ with $n=4$ or $5$) 
and many values of $z$ on the imaginary axis, $z={\rm i}r$. 
One then plots the modulus 
$|\smean{g^{0,\theta_2}_0({\rm i}r)}_{\rm evts}|$ for real, positive $r$ 
as a function of $r$, for a fixed value of $\theta_2$. 
One determines numerically the value of $r$ corresponding 
to the first minimum of this function,  $r=r_0^{\theta_2}$.
The value of $r_0^{\theta_2}$ will be used in the analysis 
of directed flow below. 
It is closely related to elliptic flow, which can be estimated 
through the following formula~\cite{Bhalerao:2003yq,Borghini:2004ke}:
\begin{equation}
\label{v2infty}
|V_2|=\mean{\frac{j_{01}}{r_0^{\theta_2}}}_{\!\!\theta_2},
\end{equation}
where $j_{01}\simeq 2.40483$ is the first root of the Bessel 
function of the first kind $J_0(x)$. 
The notation $\mean{\cdots}_{\theta_2}$ means an average 
over $\theta_2$. 
The analysis could in principle be performed with a single value of 
$\theta_2$. 
The averaging is only of practical importance: it reduces 
the statistical error on $V_2$  by a factor of $\sim 2$.

Note that Eq.~(\ref{v2infty}) yields an estimate of the absolute 
value of $V_2$, not of $V_2$ itself. 
The sign of $V_2$ is determined simultaneously with 
directed flow, as we shall see below.

\subsection{Integrated directed flow}
\label{s:IIB}

The analysis of integrated directed flow involves the generating function 
$g_\epsilon ^{\theta_1,\theta_2}(z)$. 
The angle $\theta_2$ takes the same values as in Sec.~\ref{s:IIA}. 
For a given $\theta_2$, $z$ takes only one value, 
$z={\rm i}r_0^{\theta_2}$. 

For each event, one first evaluates the generating function for 
$\epsilon=\theta_1=0$, as well as its derivative with respect to $z$:
\begin{equation}
\frac{\partial g^{0,\theta_2}_0}{\partial z}({\rm i}r_0^{\theta_2})=
g_0^{0,\theta_2}({\rm i}r_0^{\theta_2})
\sum_{j=1}^M\frac{w_{2}(j)\cos(2(\phi_j-\theta_2))}
{1+{\rm i}r_0^{\theta_2} w_{2}(j)\cos(2(\phi_j-\theta_2))}. 
\end{equation} 
One then evaluates $g_\epsilon ^{\theta_1,\theta_2}({\rm i}r_0^{\theta_2})$ 
for a single non-zero value of $\epsilon$, which is arbitrary 
but must satisfy the following condition
\begin{equation}
\label{smallepsilon}
\epsilon\ll \frac{V_2}{V_1}. 
\end{equation}
Since $V_1$ is a priori unknown, one must guess a reasonable 
value and check afterwards that the condition is satisfied. 
If $\epsilon$ is too small, numerical errors 
may arise as the procedure amounts to expanding numerically 
$g_\epsilon ^{\theta_1,\theta_2}(z)$ to order $\epsilon^2$. 
Thus we recommend to perform tests with several values of $\epsilon$, 
and to check the stability of the results, before doing the full 
analysis.  
The angle $\theta_1$ takes 5 (or more) equally spaced 
values of $\theta_1$ between 0 and $2\pi$, (i.e. 
$\theta_1=0,2\pi/n,\ldots, 2(n-1)\pi/n$ with $n\ge 5$). 
Note that the range differs from that of $\theta_2$ values.

Our estimate of $V_1$ is defined by 
\begin{equation}
\label{intv1}
(V_1)^2\,{\rm sgn}(V_2)=\mean{-\frac{8j_{01}}{\epsilon^2}
\left(r_0^{\theta_2}\right)^{\!-3}
{\rm Re}\left( \frac{\mean{
\cos(2(\theta_1-\theta_2))\,
g_\epsilon^{\theta_1,\theta_2}({\rm i}r_0^{\theta_2})}_{\!\theta_1,\rm evts}}
{\displaystyle
  \mean{\frac{\partial g^{0,\theta_2}_0}{\partial z}({\rm i}r_0^{\theta_2})}
_{\!\!\rm evts}}\right)}_{\!\!\!\raisebox{11pt}{$\scriptstyle\theta_2$}},
\end{equation}
where $\epsilon$ is an arbitrary small number satisfying 
condition~(\ref{smallepsilon}), 
$\smean{\cdots}_{\theta_1,\rm evts}$ denotes an average over $\theta_1$
and events, 
and Re is the real part of the ratio.
Finally, ${\rm sgn}(V_2)$ denotes the sign of $V_2$, 
which is determined by this analysis of $V_1$. 
The sign of $V_1$, on the other hand, is not measured 
and must be postulated, as in any other method of analysis. 

As in Sec.~\ref{s:IIA}, the averaging over $\theta_2$ in
Eqs.~(\ref{intv1}) and (\ref{diffv1}) is only for practical purposes:
it reduces statistical errors by a factor of $\sim 2$. 
In contrast, the averaging over $\theta_1$ cannot be avoided, for reasons
we shall explain below in Sec.~\ref{s:theory}.

\subsection{Differential directed flow}
\label{s:IIC}

One can then turn to the analysis of differential flow, i.e., 
the flow of particles of a given type in a definite phase-space
window, which we shall call ``protons'' for the sake of brevity.
A ``proton'' azimuth will be denoted by $\psi$, and the corresponding
differential directed flow by $v'_1$. 

The estimate of $v'_1$ involves the derivative of the 
generating function with respect to the proton weight, evaluated at 
$z={\rm i}r_0^{\theta_2}$: 
\begin{equation}
\label{partialw}
\frac{\partial g^{\theta_1,\theta_2}_{\epsilon}({\rm i}r_0^{\theta_2})}
{\partial w_1(\psi)} = 
g_\epsilon^{\theta_1,\theta_2}({\rm i}r_0^{\theta_2})
\frac{{\rm i}r_0^{\theta_2}\epsilon \cos(\psi-\theta_1)}
{1+{\rm i}r_0^{\theta_2}\epsilon w_{1}(\psi)\cos(\psi-\theta_1)
+{\rm i}r_0^{\theta_2} w_{2}(\psi)\cos(2(\psi-\theta_2))},
\end{equation} 
and is defined by 
\begin{equation}
\label{diffv1}
v'_1=\mean{-\frac{4j_{01}\,{\rm sgn}(V_2)}{V_1\epsilon^2}
\left(r_0^{\theta_2}\right)^{\!-3}
{\rm Re}\left( \frac{\displaystyle
\mean{\cos(2(\theta_1-\theta_2))\,\frac{
\partial g^{\theta_1,\theta_2}_{\epsilon}({\rm i}r_0^{\theta_2})}
{\partial w_1(\psi)}}_{\!\!\theta_1,\psi}}
{\displaystyle
  \mean{\frac{\partial g^{0,\theta_2}_0}{\partial z}({\rm i}r_0^{\theta_2})}
_{\!\!\rm evts}}\right)}_{\!\!\!\raisebox{12pt}{$\scriptstyle\theta_2$}},
\end{equation}
where $\smean{\cdots}_{\theta_1,\psi}$ denotes an average over protons
and over values of $\theta_1$. 
Note that the value of integrated directed flow $V_1$ is only required after 
averaging over protons or events, which means that the ratio between parentheses 
can be computed at the same time as that in Eq.~(\ref{intv1}). 
Moreover, the denominator of this ratio is the same as in Eq.~(\ref{intv1}), 
so that in the second pass through data, one only need compute three 
quantities for each event: this common denominator, the numerator of 
Eq.~(\ref{intv1}) (actually, a value for each angle $\theta_1$), and the 
numerator of Eq.~(\ref{diffv1}), taking into account only the ``protons.''

\subsection{Relation with the three-particle method of Ref.~\cite{Borghini:2002vp}}
\label{s:IID}

The three-particle method we proposed in Ref.~\cite{Borghini:2002vp} is 
based on averages of the type $\mean{\cos(\phi_1+\phi_2-2\phi_3)}$, 
where $\phi_1$, $\phi_2$ and $\phi_3$ are the azimuthal angles of three 
particles belonging to the same event. 
Let us explain how such three-particle averages appear in the 
present  method:
expanding $g_{\epsilon}^{\theta_1,\theta_2}(z)$ in 
Eq.~(\ref{defgenfunc}) to order $\epsilon^2 z^3$ produces 
terms of the type $\cos(\phi_1-\theta_1)\cos(\phi_2-\theta_1)
\cos(2(\phi_3-\theta_2))$. Multiplying by 
$\cos(2(\theta_1-\theta_2))$ and averaging over $\theta_1$ and 
$\theta_2$, as in Eqs.~(\ref{intv1}) and (\ref{diffv1}), one obtains 
$\frac{1}{8}\cos(\phi_1+\phi_2-2\phi_3)$, thus recovering the 
three-particle averages. 
The present method uses an expansion to order $\epsilon^2$ (see 
Sec.~\ref{s:theory}), but for a fixed value of $z$, namely 
${\rm i}r_0^{\theta_2}$: it therefore also includes the information from 
higher-order correlations, generated by higher powers of $z$, not only 
three-particle correlations.

\section{Theory}
\label{s:theory}

The general philosophy is that zeroes of generating functions 
are direct probes of collective effects~\cite{Bhalerao:2003yq}, 
while they are little sensitive to nonflow effects. 
The key property of the generating function in 
Eq.~(\ref{defgenfunc}) is its factorization property:
if a nucleus-nucleus collision can be viewed as  
the superposition of independent subsystems containing 
a few particles (for instance, independent nucleon-nucleon 
collisions), i.e., if there are no collective effects, 
the average over events 
$\smean{g^{\theta_1,\theta_2}_\epsilon(z)}_{\rm evts}$
is the product of the contributions of individual subsystems.
The zeroes of the generating function are then simply the 
zeroes of the contributions of the subsystems. 
Since the value of the generating function is $1$ for $z=0$ 
by construction, the value of $z$ for which it vanishes 
verifies $|z|\sim 1$ for dimensional reasons
(in this discussion, we assume unit weights $w_1=w_2=1$ for 
simplicity). 

If, on the other hand, there is collective flow, 
nothing prevents the zeroes from moving much closer to 
the origin, $|z|\ll 1$, and this is indeed what 
happens: the positions of the zeroes scale with the 
event multiplicity like $1/M$ (while they would be independent
of $M$ for independent subsystems). 
This relates our method to the Lee--Yang theory of phase 
transitions~\cite{Yang:be}, where the zeroes of the grand 
partition function move closer and closer to the real
axis as  the size of the system increases when, and only when, 
there is a phase transition. 

Let us now be more quantitative. 
We first evaluate the average value of 
$g_\epsilon^{\theta_1,\theta_2}(z)$ for events having exactly 
the same reaction-plane orientation $\Phi_R$ (denoting such an average 
by $\smean{\cdots|\Phi_R}$). 
Taking the logarithm of Eq.~(\ref{defgenfunc}) and expanding to 
order $z$, we obtain
\begin{eqnarray}
\label{gphiR}
\ln\mean{g_\epsilon^{\theta_1,\theta_2}(z)|\Phi_R}
&\simeq&z\epsilon 
\mean{\sum_{j=1}^M w_{1}(j)\cos(\phi_j-\theta_1)|\Phi_R}
+z\mean{\sum_{j=1}^M w_{2}(j)\cos(2(\phi_j-\theta_2))|\Phi_R}\cr
&\simeq&z\epsilon  V_1\cos(\Phi_R-\theta_1)+
z V_2\cos(2(\Phi_R-\theta_2)),
\end{eqnarray}
where we have used Eqs.~(\ref{defV2}) and (\ref{defV1}) 
and assumed symmetry with respect to the reaction plane, 
which implies 
$\smean{\sum_j w_2(j)\sin(2(\phi_j-\Phi_R))}_{\rm evts}=
\smean{\sum_j w_1(j)\sin(\phi_j-\Phi_R)}_{\rm evts}=0$.
Terms of order $z^2$ and higher in the expansion are responsible for 
systematic errors that will be estimated in Sec.~\ref{s:systerrors}. 
The average over events is eventually obtained by averaging 
over $\Phi_R$, which is randomly distributed:
\begin{equation}
\label{avphiR}
\mean{g_\epsilon^{\theta_1,\theta_2}(z)}_{\rm evts}
=\int_{0}^{2\pi}\!\mean{g_\epsilon^{\theta_1,\theta_2}(z)|\Phi_R}
\frac{{\rm d}\Phi_R}{2\pi}.
\end{equation}

For $\epsilon=0$, substituting the estimate~(\ref{gphiR}) into 
Eq.~(\ref{avphiR}), one obtains
\begin{equation}
\label{oldg0}
\mean{g_0^{0,\theta_2}(z)}_{\!\rm evts}=I_0(|V_2| z),
\end{equation}
where $I_0$ is the modified Bessel function of order 0. Its zeroes 
lie on the imaginary axis, the first one being located 
at ${\rm i} j_{01}$. 
This explains how $|V_2|$ is obtained 
in Sec.~\ref{s:IIA}, Eq.~(\ref{v2infty}). 
Since zeroes are expected to be on the imaginary axis except 
for irrelevant statistical fluctuations and/or detector
effects~\cite{Bhalerao:2003yq}, we suggest to look for the position of the first 
minimum of $|\smean{g_0^{0,\theta_2}({\rm i}r)}_{\rm evts}|$ on the upper
imaginary axis, denoted by $z={\rm i}r_0^{\theta_2}$ in Sec.~\ref{s:IIA}, 
rather than that of the first zero of  
$\smean{g_0^{0,\theta_2}(z)}_{\rm evts}$ in the complex plane. 

In order to obtain an estimate of $V_1$, we shall now study how the 
first zero of $\smean{g_\epsilon^{\theta_1,\theta_2}(z)}_{\rm evts}$, 
which we denote by $z_{\epsilon}$, varies for small $\epsilon$.
For small enough $\epsilon$, we may write
\begin{equation}
\label{zepsilon}
z_\epsilon-z_0=
-\frac{\mean{g_\epsilon^{\theta_1,\theta_2}(z_0)}_{\rm evts}}
{\mean{\frac{\displaystyle\partial g^{0,\theta_2}_0}
{\displaystyle\partial z}(z_0)
}_{\!\!\rm evts}}, 
\end{equation}
where $z_0$ is the value of $z_\epsilon$ for $\epsilon=0$. 
According to  our general philosophy, the value of $z_\epsilon$
is directly related to collective flow, and little sensitive
to nonflow effects. Then, this is also true of the 
right-hand side (r.h.s.) of Eq.~(\ref{zepsilon}), which is experimentally  
measurable. 
Following the discussion above, we replace the exact zero $z_0$ 
by ${\rm i}r_0^{\theta_2}$.
To relate the quantity in Eq.~(\ref{zepsilon}) to anisotropic flow, 
we use our theoretical estimate of the generating function, 
defined by Eqs.~(\ref{gphiR}) and (\ref{avphiR}). 
The denominator is simply obtained by differentiating
Eq.~(\ref{oldg0}):
\begin{equation}
\label{denomflow}
\mean{\frac{\partial g^{0,\theta_2}_0}{\partial z}(z_0)}_{\!\!\!\rm evts}=
|V_2|\, I_1(|V_2| z_0).
\end{equation} 
The numerator is obtained by substituting Eq.~(\ref{gphiR}) into 
(\ref{avphiR}), and expanding to order $\epsilon^2$:
\begin{equation}
\label{numflow}
\mean{g_\epsilon^{\theta_1,\theta_2}(z_0)}_{\rm evts}
=\frac{\epsilon^2}{4}z_0^2\, I_1(V_2 z_0)\,V_1^2\cos(2(\theta_1-\theta_2)).
\end{equation}
Since $I_1$ is an odd function, we may write 
$I_1(V_2 z_0)=I_1(|V_2| z_0)\,{\rm sgn}(V_2)$. 
Putting together the last two equations, replacing $z_0$ with
${\rm i}r_0^{\theta_2}$ and $|V_2|$ with its estimate 
$j_{01}/r_0^{\theta_2}$ (see Eq.~(\ref{v2infty})), we obtain
\begin{equation}
V_1^2\,{\rm sgn}(V_2)\cos(2(\theta_1-\theta_2))=
-\frac{4 j_{01}}{\epsilon^2}\left(r_0^{\theta_2}\right)^{\!-3}
\frac{\mean{g_\epsilon^{\theta_1,\theta_2}({\rm i}r_0^{\theta_2})}_{\!\rm evts}}
{\mean{\frac{\displaystyle\partial g^{0,\theta_2}_0}{\displaystyle\partial z}
({\rm i}r_0^{\theta_2})
}_{\!\!\rm evts}}.  
\end{equation}
This yields in principle an estimate of $V_1$ for each value of $\theta_1$ and
$\theta_2$. However, 
the expansion of 
$\smean{g_\epsilon^{\theta_1,\theta_2}({\rm i}r_0^{\theta_2})}_{\rm evts}$
in powers of $\epsilon$ generally yields a non-vanishing term 
of order $\epsilon$ due to statistical fluctuations. 
Multiplying the previous equation by 
$\cos(2(\theta_1-\theta_2))$, and averaging over 5 or more equally 
spaced values of $\theta_1$, one eliminates this term. 
Averaging over $\theta_2$, one obtains our final estimate, 
Eq.~(\ref{intv1}). 

We can now turn to differential directed flow. 
For this purpose, let us study how the zero of 
$g_\epsilon^{\theta_1,\theta_2}(z)$ moves when the weights 
$w_1$ of all protons are shifted by some small quantity $\delta w_1$:
\begin{equation}
z_{\delta w}-z_0=-\frac{
\mean{\frac{\displaystyle \partial g_\epsilon^{\theta_1,\theta_2}(z_0)}
{\displaystyle\partial w_1(\psi)}}_{\!\psi}}
{\mean{\frac{\displaystyle\partial g^{0,\theta_2}_0}
{\displaystyle\partial z}(z_0)
}_{\!\!\rm evts}}\delta w_1. 
\end{equation} 
The value of $z_{\delta w}$ is directly related to collective effects. 
Following the same procedure as for integrated directed flow,
we estimate the value of the r.h.s. when flow is present. 
To evaluate the numerator, we first compute the 
average for a fixed $\Phi_R$. Using Eq.~(\ref{partialw}),
assuming that the proton and the other particles are uncorrelated 
for fixed $\Phi_R$ (i.e., that the correlation between the proton 
and the other particles is only due to flow), and neglecting the 
contribution of the proton to the generating function, we obtain 
\begin{eqnarray}
\mean{\frac{\partial g^{\theta_1,\theta_2}_{\epsilon}(z_0)}
{\partial w_1(\psi)}|\Phi_R}
&\simeq &\epsilon z_0\mean{\cos(\psi-\theta_1)|\Phi_R}
\mean{g_\epsilon^{\theta_1,\theta_2}(z_0)|\Phi_R}\cr
&=&\epsilon z_0 v'_1\cos(\Phi_R-\theta_1)
\exp\left[z_0\epsilon  V_1\cos(\Phi_R-\theta_1)+
z_0 V_2\cos(2(\Phi_R-\theta_2))\right],
\end{eqnarray}
where we have used Eq.~(\ref{gphiR}), the definition of 
$v'_1=\mean{\cos(\psi-\Phi_R)}$, and assumed symmetry with 
respect to the reaction plane, $\mean{\sin(\psi-\Phi_R)}=0$. 
We then expand the exponential to order $\epsilon$ and
integrate over $\Phi_R$: 
\begin{equation}
\label{numdiff}
\mean{\frac{\partial g^{\theta_1,\theta_2}_{\epsilon}(z_0)}
{\partial w_1(\psi)}}
=\frac{\epsilon^2}{2}z_0^2\, v'_1V_1\, 
I_1(V_2 z_0)\,\cos(2(\theta_1-\theta_2)).
\end{equation}
Using Eq.~(\ref{denomflow}), and replacing $z_0$ with 
${\rm i}r_0^{\theta_2}$, we obtain 
\begin{equation}
v'_1 V_1\, {\rm sgn}(V_2)\cos(2(\theta_1-\theta_2))=-
\frac{2 j_{01}}{\epsilon^2}\left(r_0^{\theta_2}\right)^{\!-3}
\frac{
\mean{\frac{\displaystyle \partial g_\epsilon^{\theta_1,\theta_2}
({\rm i}r_0^{\theta_2})}
{\displaystyle\partial w_1(\psi)}}_{\!\psi}}
{\mean{\frac{\displaystyle\partial g^{0,\theta_2}_0}
{\displaystyle\partial z}({\rm i}r_0^{\theta_2})
}_{\!\!\rm evts}}.
\end{equation}
Multiplying both sides of this equation by $\cos(2(\theta_1-\theta_2))$
and averaging over $\theta_1$, one eventually obtains Eq.~(\ref{diffv1}).

For the sake of consistency, one must recover the integrated 
flow $V_1$ by integrating the differential flow $v'_1$ over phase
space. 
Our estimates (\ref{intv1}) and (\ref{diffv1}) satisfy this 
sum rule, provided $\epsilon$ is small enough. In order to 
prove this, we weight  
Eq.~(\ref{diffv1}) with $w_1(\psi)$ and integrate over 
phase space. The following quantity appears 
\begin{equation}
\sum_{j=1}^M w_1(j)\frac{\partial g^{\theta_1,\theta_2}_{\epsilon}({\rm i}r_0^{\theta_2})}
{\partial w_1(j)}=
\epsilon
\frac{\partial g^{\theta_1,\theta_2}_{\epsilon}({\rm i}r_0^{\theta_2})}
{\partial\epsilon}. 
\end{equation}
Next, we use the fact that
$g^{\theta_1,\theta_2}_{\epsilon}({\rm i}r_0^{\theta_2})$ scales like 
$\epsilon^2$ for small $\epsilon$, hence 
\begin{equation}
\epsilon
\frac{\partial g^{\theta_1,\theta_2}_{\epsilon}({\rm i}r_0^{\theta_2})}
{\partial\epsilon}=
2g^{\theta_1,\theta_2}_{\epsilon}({\rm i}r_0^{\theta_2}).
\end{equation}
This completes the proof.

\section{Errors}
\label{s:errors}

The main motivation for analyzing collective flow with 
multiparticle correlations is that one thereby reduces 
spurious, ``nonflow'' effects. The magnitude of errors due 
to residual nonflow effects is estimated in Sec.~\ref{s:systerrors}. 
The price to pay for this greater reliability is 
an increase in statistical errors, which are evaluated 
in Sec.~\ref{s:staterrors}. 
Finally, azimuthal asymmetries in the detector acceptance 
are a potential source of bias in the analysis. They are 
carefully studied in Sec.~\ref{s:acceptance}.

\subsection{Systematic errors from nonflow effects}
\label{s:systerrors}

There are several tricks to check experimentally whether or 
not the analysis is biased by nonflow effects. 
A first one is to perform the analysis twice, using different weights
in Eq.~(\ref{defgenfunc}), 
e.g., one analysis with $w_2=p_t$ and another with $w_2=1$, 
or redoing the analysis with zero weights for the 
particles in one hemisphere. 
The final differential flow results should be independent of 
the weights, while no such independence is expected for nonflow 
effects. However, 
using non-optimal weights increases statistical errors, 
sometimes by large amounts, so that this is doable in practice 
only if statistical errors are not a limitation. 
A second trick is to check that the final results look 
``reasonable:'' this means in particular that they must be consistent 
with the known symmetries of the system, i.e., $v_1(-y)=-v_1(y)$, 
$v_2(-y)=v_2(y)$ for a symmetric collision. 
Nonflow effects can destroy these symmetries: for instance, effects of 
total momentum conservation may contaminate the measurement of $v_1$, 
yielding an estimate that does not vanish near 
midrapidity~\cite{Borghini:2002mv}. 
This may however occur only when the weights in Eq.~(\ref{defgenfunc}) 
are not symmetric themselves (i.e., $w_1(-y)\not=-w_1(y)$, 
$w_2(-y)\not= w_2(y)$). With symmetric weights, the fact that 
$v_1$ vanishes at midrapidity is not an indication that it is not 
biased by nonflow effects. 

In this paper, we are interested in directed flow at ultrarelativistic 
energies, which is most often at the border of observability, so 
that it is not always possible to use the above tricks. 
It is therefore important to derive estimates of the magnitude of nonflow 
effects using purely theoretical arguments. There is of course 
no way to derive quantitative estimates, due to the variety of the 
physical effects involved~\cite{Borghini:2000cm}. 
In most cases, one can at best rely on orders of magnitude 
and simple scaling rules~\cite{Borghini:2000sa}, which 
are derived below. 
More quantitative statements can be made concerning momentum 
conservation, a well-known bias in analyses of directed 
flow~\cite{Borghini:2002mv,Danielewicz:1988in,Ogilvie:1989}.  

With standard methods of flow 
analysis~\cite{Danielewicz:hn,Poskanzer:1998yz} and also, 
to a lesser degree, with cumulants~\cite{Borghini:2000sa},
the analysis may yield a non-zero value of $v_1$ and $v_2$ even 
when they are in fact zero, i.e., even if only ``nonflow'' effects 
are present~\cite{Kovchegov:2002nf}. 
This is not the case with Lee--Yang zeroes, where a non-vanishing
result beyond statistical errors can be considered a clear 
signal of anisotropic flow. 

However, the interference of collective flow and 
nonflow effects may produce a small relative error, 
which  is the sum of two terms:
\begin{equation}
\label{systerror}
\frac{(\delta v_1)_{\rm nonflow}}{v_1}
={\cal O}\!\left(\frac{v_2}{M v_1^2}\right) +
{\cal O}\!\left(\frac{1}{M v_2}\right). 
\end{equation}
Let us explain how such terms arise. 
Our determination of directed flow with the 
present method involves three-particle averages
such as $\mean{\cos(\phi_1+\phi_2-2\phi_3)}$
(see Sec.~\ref{s:IID}).\footnote{One can check that the systematic 
errors due to higher-order correlations are at most of the same order 
as those due to three-particle correlations.}
The contribution of flow to this average is $v_1^2 v_2$. 
Let us now evaluate the contribution of nonflow effects. 
As a simple model, assume that the $M$ particles 
detected in an event are emitted in $M/2$ collinear pairs 
(``jet-like correlations'').  
There is a probability $1/(M-1)\simeq 1/M$ that $\phi_2=\phi_1$, 
in which case the three-particle average becomes 
$\mean{\cos 2(\phi_1-\phi_3)}=v_2^2$. This gives the first 
term in Eq.~(\ref{systerror}). 
Similarly, there is a probability of order $1/M$ that 
$\phi_1=\phi_3$, in which case the 
three-particle average reads
$\mean{\cos (\phi_1-\phi_2)}=v_1^2$; 
this yields the second term in Eq.~(\ref{systerror}). 
Since the present method is expected to be useful essentially when 
$v_1<v_2$, this term is subleading. 
Finally, one can check explicitly that momentum conservation 
does not contribute to the error terms in Eq.~(\ref{systerror}). 

The order of magnitude of the systematic error, Eq.~(\ref{systerror}), is more 
general than suggested by this toy model of nonflow correlations. 
It can be proven more rigorously by pushing the expansion of the logarithm of 
the generating function, Eq.~(\ref{gphiR}), to order $z^2$, and studying the 
influence of the additional terms on the flow estimate, much in the same way 
as in Ref.~\cite{Bhalerao:2003yq}, where the interference between flow and 
nonflow effects was also considered (see the discussion following Eq.~(48) in 
the reference). 
Of course, Eq.~(\ref{systerror}) is but a scaling law, and each term involves 
an unknown numerical coefficient. 
One can try to estimate these coefficients experimentally by studying the 
difference between flow estimates from various methods, using situations where 
this difference is believed to be dominated by nonflow effects. 
This strategy was applied by the STAR Collaboration~\cite{Adams:2003zg} in 
their estimates of systematic errors on $v_4$ and $v_1$. 

With the standard, event-plane method~\cite{Danielewicz:hn,Poskanzer:1998yz}, 
the error due to nonflow effects is much larger:
\begin{equation}
\label{syststandard}
\frac{(\delta v_1)_{\rm nonflow}}{v_1}
={\cal O}\!\left(\frac{1}{M v_1^2}\right).
\end{equation}
Comparing with Eq.~(\ref{systerror}), our method reduces 
the nonflow error by a factor of $1/v_2$, i.e., typically 
15 to 20 at RHIC.  

The errors with the present method, Eq.~(\ref{systerror}), 
also apply to the three-particle cumulant method, but 
they were not mentioned in Ref.~\cite{Borghini:2002vp}. 
The cumulant method has an additional error term, 
coming from the ``pure nonflow'' contribution to the 
three-particle average $\mean{\cos(\phi_1+\phi_2-2\phi_3)}$, 
of order $1/M^2$. 
This gives an additional error term in Eq.~(\ref{systerror}) 
\begin{equation}
\label{systerror3}
\frac{(\delta v_1)_{\rm nonflow}}{v_1}
={\cal O}\!\left(\frac{1}{M^2 v_2 v_1^2}\right). 
\end{equation}
This contribution, which was the only error 
term mentioned in Ref.~\cite{Borghini:2002vp},
disappears with Lee--Yang zeroes. 
In practice at ultrarelativistic energies $Mv_2^2\gtrsim 1$, so that the 
additional term is typically of the same order of magnitude as 
the first term of Eq.~(\ref{systerror}). In addition, 
momentum conservation produces an error of this order, 
so that the Lee--Yang zeroes method does a better job
in eliminating effects of momentum conservation.
Finally, the three-particle method requires to estimate $v_2$ 
independently. If $v_2$ is estimated from two-particle 
cumulants or by the event-plane method, the resulting 
error on $v_1$ is 
$(\delta v_1)_{\rm nonflow}/v_1={\cal O}(1/M v_2^2)$, 
which may dominate over both terms in Eq.~(\ref{systerror}).

\subsection{Statistical errors}
\label{s:staterrors}

Statistical uncertainties on $V_1$ and $v'_1$ naturally involve 
both directed and elliptic flows. 
Both errors turn out to scale like the statistical error on 
the differential elliptic flow $v'_2$ determined 
from Lee--Yang zeroes~\cite{Bhalerao:2003yq}, which is natural since 
we use elliptic flow as a reference. 
The statistical error on elliptic flow involves the 
associated {\it resolution parameter\/} 
\begin{equation}
\label{chi2}
\chi_2\equiv\frac{V_2}{\sigma_2},
\end{equation}
where $\sigma_2$ is a measure of event-by-event fluctuations.
Provided nonflow effects are not too large, $\sigma_2$ can be obtained 
experimentally by the following formula:
\begin{equation}
\label{sigma2}
\sigma_2^2\simeq \mean{\sum_{j=1}^M w_2(j)^2}_{\!\!\rm evts}.
\end{equation}
More accurate determinations are discussed in Ref.~\cite{Bhalerao:2003yq}. 
Once $\chi_2$ has been determined, the statistical error on $v'_2$ 
is~\cite{Bhalerao:2003yq}:
\begin{equation}
\label{deltav2}
(\delta v'_2)^2=\frac{1}{4\Np J_1(j_{01})^2}
\mean{
\exp\left(\frac{j_{01}^2}{2\chi_2^2}\cos 2\theta_2\right)
J_0(2 j_{01}\sin\theta_2)
-\exp\left(-\frac{j_{01}^2}{2\chi_2^2}\cos 2\theta_2\right)
J_0(2 j_{01}\cos\theta_2)}_{\!\theta_2}, 
\end{equation} 
where $\Np$ is the number of ``protons'' in the phase-space region 
under study. 
For large $\chi_2$, this yields 
$\delta v'_2=1/\sqrt{2\Np}$, which is the expected statistical 
error when the reaction plane is exactly known. 
When $\chi_2$ becomes significantly smaller than unity, on the 
other hand, the statistical error on $v'_2$ increases exponentially 
as $\delta v'_2\propto \exp(2.9/\chi_2^2)$, and the method cannot
be applied~\cite{Bhalerao:2003yq}. 

Let us now discuss statistical errors on the directed flow estimates. 
These errors involve two parameters $\sigma_1$ and 
$\chi_1$ defined as in Eqs.~(\ref{chi2}) and (\ref{sigma2}), 
namely, $\sigma_1^2\simeq\smean{\sum_j w_j^2}_{\rm evts}$ 
and $\chi_1=V_1/\sigma_1$. 
The latter is the resolution parameter associated with directed flow.%
\footnote{Please note that the resolution parameter $\chi$ as defined 
in Ref.~\cite{Poskanzer:1998yz} is larger by a factor of $\sqrt{2}$} 
In order to obtain the square of the statistical error on 
$V_1^2{\rm sgn}(V_2)$, one then does the following substitution 
in Eq.~(\ref{deltav2}):
\begin{equation}
\frac{1}{\Np}\to \frac{2\,\sigma_1^4(1+\chi_1^2)}{N_{\rm evts}}.
\end{equation}
For large $\chi_1$ and $\chi_2$, this yields 
$\delta V_1=\sigma_1/\sqrt{2 N_{\rm evts}}$, which is the 
expected value of the statistical error if the reaction plane 
is exactly known. 

Finally, the statistical error on the differential directed flow 
$v'_1$ is simply related to the error on $v'_2$:
\begin{equation}
\label{statervp1}
(\delta v'_1)^2=\frac{1+\chi_1^2}{\chi_1^2}(\delta v'_2)^2.
\end{equation}
One notes that the statistical error on directed flow obtained 
with this method is always larger than the statistical error 
on elliptic flow in absolute value.  
When $\chi_2$ is large enough, $\delta v'_1$ is the same as 
with the event-plane analysis.  

When using 3-particle cumulants~\cite{Borghini:2002vp,Adams:2003zg},
the statistical error is given by a similar formula, where 
$\delta v'_2$ is the error on elliptic flow obtained with 
the event-plane method (or by 2-particle correlations). 
Going from the 3-particle method to the present method, 
the increase in statistical errors will be the same as 
when going from the standard method to Lee--Yang zeroes in
the analysis of elliptic flow. 
For a semi-central Au-Au collision at RHIC analyzed with the 
STAR detector, the resolution parameter $\chi_2$ is at least 1, 
which means a factor of at most 2 increase in statistical 
errors~\cite{Bhalerao:2003yq}. 
This increase is compensated by the smaller error from nonflow effects.

\subsection{Azimuthal asymmetries in the detector acceptance}
\label{s:acceptance}

A nice feature of all flow analyses based on cumulants or Lee--Yang 
zeroes (which amount to cumulants of many-particle 
correlations~\cite{Bhalerao:2003yq}) is that they automatically 
eliminate most effects of azimuthal asymmetries 
in the detector acceptance: 
more precisely, such asymmetries cannot produce a signal 
by themselves, even if the detector has a very partial 
azimuthal coverage. This is not the case with the event-plane 
analysis, where several flattening procedures are
required~\cite{Poskanzer:1998yz}, and with two-particle 
correlation methods~\cite{Wang:1991qh,PHENIX}, which use 
mixed events to correct for acceptance effects. 

Acceptance asymmetries generally have two effects with 
cumulants or Lee--Yang zeroes~\cite{Borghini:2000sa,Bhalerao:2003yq}: 
1) the flow given by 
the analysis differs from the true flow by some factor,
which can be computed analytically once the acceptance 
profile is known; 
2) different harmonics interfere, so 
that a measurement of $v_1$ is generally biased by $v_2$ 
and vice-versa. These interference terms can also be calculated 
analytically.

Such interference terms between $v_1$ and $v_2$ are also 
present here, and they turn out to be a rather serious problem. 
Indeed, as an explicit calculation will show, acceptance 
asymmetries produce an additional term proportional to 
$v_2$ in the left-hand side (l.h.s.) of Eq.~(\ref{intv1}): even 
if there is only elliptic flow in the system, the analysis 
then yields a spurious directed flow $v_1\propto\sqrt{v_2}$. 
Since we are typically interested in a situation where 
$v_2$ is large and $v_1$ is small, such an interference 
is most unwelcome, and it must be thoroughly studied. 

In order to focus on the interference with $v_2$, we assume 
that only elliptic flow is present in the system, $v_1=0$, and 
we evaluate the spurious directed flow. 
The following study is limited to the integrated flow $V_1$. 
For a given reaction plane $\Phi_R$, the azimuthal distribution 
of a particle of type $j$ is, before detection, 
\begin{equation}
\frac{{\rm d}N}{{\rm d}\phi_j}\propto 1+2 v_2(j)\cos(2(\phi_j-\Phi_R)). 
\end{equation}
With this distribution, and with a non-symmetric detector, 
the first term in Eq.~(\ref{gphiR}) becomes 
\begin{equation}
\mean{\sum_{j=1}^M w_{1}(j)\cos(\phi_j-\theta_1)|\Phi_R}=
{\rm Re}\left[ V_0^{\rm acc}{\rm e}^{-{\rm i}\theta_1}
+V_2^{\rm acc} {\rm e}^{{\rm i}(2\Phi_R-\theta_1)}
+V_2^{\prime\,\rm acc} {\rm e}^{{\rm i}(-2\Phi_R-\theta_1)}\right],
\end{equation}
where $V_0^{\rm acc}$, $V_2^{\rm acc}$ and $V_2^{\prime\,\rm acc}$
are complex coefficients defined by 
\begin{eqnarray}
V_0^{\rm acc}&=&
  \mean{\sum_{j=1}^M w_{1}(j)\,{\rm e}^{{\rm i}\phi_j}}\cr
V_2^{\rm acc}&=&
  \mean{\sum_{j=1}^M w_{1}(j)\,v_2(j)\,{\rm e}^{-{\rm i}\phi_j}}\cr
V_2^{\prime\,\rm acc}&=&
  \mean{\sum_{j=1}^M w_{1}(j)\,v_2(j)\,{\rm e}^{3{\rm i}\phi_j}},
\end{eqnarray}
with the averages taken over detected particles. 
For a symmetric acceptance, the three coefficients vanish. 
With these values, one can compute the generating function 
$\mean{g_\epsilon^{\theta_1,\theta_2}(z)}$ to order $\epsilon^2$
following essentially the same steps as in Sec.~\ref{s:theory}. 
Let us give the final result, skipping tedious intermediate
calculations: to leading order in acceptance asymmetries, 
one must replace $(V_1)^2$ with ${\rm Re}\,[2 V_0^{\rm acc}V_2^{\rm acc}]$ 
in the l.h.s.\ of Eq.~(\ref{intv1}). 

For moderate inhomogeneities in the detector acceptance, the interference
of elliptic flow on the estimate of directed flow remains under control. 
Thus, assuming an elliptic flow value $v_2=0.06$, variations of 20\% in 
the detector efficiency will result in a spurious $v_1$ of at most 0.01. 
Using appropriate weights (e.g., weighting with the inverse of the detector 
efficiency profile) would significantly decrease this spurious directed 
flow value. 
However, we fear that working with a detector with incomplete azimuthal 
coverage might prove prohibitive. 

Finally, note that the three-particle method is affected by 
acceptance asymmetries in essentially the same way, but 
the corresponding interference term was unfortunately omitted 
in Ref.~\cite{Borghini:2002vp}.

\section{Conclusions and perspectives}

In the foregoing, we have introduced a new method to analyze directed flow
$v_1$ in cases where it is too small to be reliably obtained with standard
two-particle methods without contamination from nonflow effects, and also too
small to be determined by more recent methods as cumulants or Lee--Yang 
zeroes, which would both give large statistical uncertainties.
Moreover, the method relies on the implicit assumption that elliptic flow is
reasonably large, as when using three-particle 
cumulants~\cite{Borghini:2002vp}.
With respect to the latter, the new method is easier to implement: instead of
interpolating successive derivatives of a function, one need just determine
the position $r_0$ of the first minimum of a function, and then compute the
values at $r_0$ of the three quantities that appear in the numerators and 
denominator of Eqs.~(\ref{intv1}) and (\ref{diffv1}).
In addition, the method is conceptually more elegant, since it relies on the
deep relation there exists between the behaviour of the zeroes of generating
functions and the presence of collective effects in the systems these 
functions describe~\cite{Bhalerao:2003yq}.

We have also carefully estimated both systematic errors due to nonflow effects
and statistical uncertainties arising from finite available statistics.
We may use these estimates to discuss the applicability of the method to 
present and future heavy-ion experiments. 
At RHIC, the three-particle cumulant method was already successfully 
applied~\cite{Adams:2003zg}. 
In mid-central collisions, statistical uncertainties should be at most a
factor 2 larger with the new method, but with the high statistics of Run 4, 
this should not be a problem.
On the other hand, systematic errors will be reduced:
the error term (\ref{systerror3}), which arises in particular 
from momentum conservation, disappears with the new method. 
The method could be useful for new measurements of directed flow at 
the CERN SPS, although this would require detectors with a larger coverage 
than what single experiments had in the past runs.

Finally, one can expect that the present method will allow measurements of
directed flow at LHC, if any, using muons from decay pions seen in the 
ALICE spectrometer at forward rapidities ($2.5<\eta<4$) or hits in the 
CMS very forward hadronic calorimeter ($3<\eta<5$). 
Since $v_2$ at LHC is expected to be at least as large as at RHIC,  
and the multiplicity will be higher, elliptic flow will be analyzed
with an excellent resolution. Then, the statistical errors 
with our new method will be barely larger than with the standard 
event-plane method: statistics will not be a limitation down 
to values of $v_1$ of a fraction of a percent. 
Systematic errors from nonflow effects may be a more severe problem, 
although our method minimizes their magnitude. 
Comparing Eq.~(\ref{statervp1}), with $\chi_1\sim v_1\sqrt{M}$, 
and the first term of Eq.~(\ref{systerror}), one sees that as 
$v_1$ decreases, systematic errors tend to become larger than 
statistical errors. One may reasonably hope, however, that the 
huge particle multiplicity $M$ expected at LHC will compensate 
for the smaller value of $v_1$, and that $v_1$ will eventually 
be observed.


\end{document}